\documentclass{aa}
\usepackage{graphicx}
\begin{document}
   \title{Nonuniform viscosity in the solar nebula and large masses of Jupiter and Saturn}

   \author{Liping Jin
          \inst{1,}\inst{2}\fnmsep\thanks{Early ideas on this research began at Clemson Univ.
          The author is also
          a visiting scholar at Univ of North Carolina at Chapel Hill.}
          }


   \institute{College of Physics, Jilin University, 2519 Jie Fang Rd, Changchun, Jilin 130021, P.R.China\\
              \email{jinlp@jlu.edu.cn}
         \and
             Dept. of Physics and Astronomy, Clemson Univ., Clemson SC 29634, USA\\
             }

   \date{Received June 1, 2004; accepted June 24, 2004}

   \abstract{
I report a novel theory that {\it nonuniform} viscous frictional
force in the solar nebula accounts for the largest mass of Jupiter
and Saturn and their largest amount of H and He among the planets,
two outstanding facts that are unsolved puzzles in our
understanding of origin of the Solar System. It is shown that the
nebula model of uniform viscosity does not match the present
planet masses. By studying current known viscosity mechanisms, I
show that viscosity is more efficient in the inner region inside
Mercury and the outer region outside Jupiter-Saturn than the
intermediate region. The more efficient viscosity drives faster
radial inflow of material during the nebula evolution. Because the
inflow in the outer region is faster than the intermediate region,
the material tends to accumulate in Jupiter-Saturn region which is
between the outer and intermediate region. It is demonstrated that
the gas trapping time of Jovian planets is longer than the inflow
time in the outer region. Therefore the gas already flows to
Jupiter-Saturn region before Uranus and Neptune can capture
significant gas. But the inflow in the Jupiter-Saturn region is so
slow that they can capture large amount of gas before the gas can
flow further inward. Hence they have larger masses with larger H
and He content than Uranus and Neptune. I also extend the
discussion to the masses of the terrestrial planets, especially
low mass of Mercury. The advantages of this theory are discussed.
   }

   \maketitle
%

\section{Introduction}

The most widely accepted theory of the origin of the Solar System
is the solar nebula hypothesis. Although the theory is, in
general, compatible with most of observational facts from the
Solar System and star forming regions (Lissauer 1993), a model in
which all details are right is not produced yet. Two of the
outstanding facts that any theory of the origin of the Solar
System must explain are planet masses and their bulk compositions
(Lissauer 1993). Jovian planets (Jupiter, Saturn, Uranus, and
Neptune) have much larger masses than terrestrial planets
(Mercury, Venus, Earth, and Mars). Furthermore Jupiter and Saturn
are by far the most massive among the Jovian planets. The
terrestrial planets are composed of rocky material while the
Jovian planets contain both heavy elements and H, He. The
abundance of H and He decreases outwards from Jupiter. In this
letter, I take the widely used scenario from both the origin of
the Solar System and star forming regions. And in that context, I
discuss the impact of $nonuniform$ viscosity mechanism on the
planet masses and their bulk compositions.

The star formation theory suggests that stars are formed from
gravitational collapses of molecular cloud cores. Due to the
existence of angular momentum, not all material collapses directly
into the center to form stars, rather systems of star+disk are
formed (the solar nebula is such a disk). It is well known that the
internal viscous frictional force in the disk transports the angular
momentum outward and the bulk of the disk material diffuses inwards
onto the protostar. This naturally explains the angular momentum
distribution of the Solar System. This viscous force is often
referred to as the angular momentum transport (AMT). Planets are
formed in the nebula (Lissauer 1993; Wuchterl et al. 2000). As the
material infall slows and ceases, the nebula becomes cool enough
that condensates may form. The grains grow via collisions into solid
bodies known as planetesimals and sediment towards the midplane of
the nebula. These planetesimals then grow into the terrestrial
planets and cores of the Jovian planets. The Jovian planets form in
the outer part which cools first and is colder than the inner part.
Therefore a head start plus ice condensation providing sufficient
mass enable the Jovian planets to trap the gases in the nebula
before they are dispersed. This accounts for the large masses of the
Jovian planets and their abundances of H and He while the
terrestrial planets are composed of rocky material and have small
masses.

Although, based on the current nebula theory and planet formation
theory, the interpretation for the difference between the Jovian
and terrestrial planets is accepted, the reason why Uranus and
Neptune have smaller masses and less H and He than Jupiter and
Saturn is still a puzzle and no theory is widely accepted. 1) The
photoevaporation theory (Hollenbach et al. 2000) brings an extra
physical process and uncertainties with it, e.g. the assumed
photoionization rate exceeds the current solar ultraviolet output
by two orders of magnitude. 2) Hydrodynamic accretion theory
(Wuchterl et al. 2000; Wuchterl 1995) requires a nebula density
that may not be realistic. 3) It is suggested (Wuchterl et al.
2000; Pollack et al. 1996) that the cores of Uranus and Neptune
grow more slowly than those of Jupiter and Saturn and do not
obtain enough mass to accrete large amount of gas before the gas
is dispersed. This scenario requires an effective dispersal
process to disperse the gas in Uranus-Neptune region before they
accrete large amount of the gas. Such a dispersal process is
suggested here. In this letter, I will demonstrate that nonuniform
viscous force in the nebula offers a natural solution to this
puzzle. I also show advantages of the theory suggested here at the
end.

\section{Inconsistency of the nebula model of uniform viscosity with the present planet masses}

The present mass distribution of the planets should give us some
idea of the surface density of the nebula. Since the gas is
dispersed later, the mass distribution of heavy elements of the
planets should reflect the surface density. The well-known minimum
mass solar nebula model (Hayashi 1981) is constructed as follows:
the material in each planet is recovered to the solar composition
and spread over an annulus reaching halfway to the orbits of its
neighbors. By using the ``annulus" approach, the surface density
$\Sigma$ from any model can be compared to the masses of the
planets. Many nebula models have been built based on a $constant$
$\alpha$ viscosity (uniform AMT) where the viscous stress is scaled
with pressure $P$ as $\alpha P$. For example, the similarity
solution by Hartmann et al. (1998) shows that $\Sigma$ varies as
$\sim r^{-1}$ at small radii and falls sharply at large distances
(where $r$ is the heliocentric radius). Notice that $\Sigma$
decreases outward with $r$. I list, in Table 1 (in units of the
earth mass, $M_\oplus$), $M_\alpha$, calculated heavy element masses
with $\Sigma \sim r^{-1}$ and $M_h$, measured masses of the
terrestrial planets or masses of heavy elements of the Jovian
planets inferred from current planet model (Guillot 1999).
$M_\alpha$ is scaled with the heavy element mass of Uranus. From
this table, by comparing heavy element masses of the planets with
those obtained from $\Sigma \sim r^{-1}$, I discover that {\it the
nebula model of constant $\alpha$ (uniform AMT) does not match the
planet masses} and {\it Jupiter and Saturn masses are significantly
enhanced}. Notice that by ``the mass enhancement" throughout this
letter, I mean the enhancement compared with the nebula model of
$constant$ $\alpha$. The terrestrial planets also have some
enhancement except the famous Mars drop and low Mercury mass. I will
discuss these together later.

   \begin{table}
   \centering
      \caption[]{Planet masses of heavy elements.}
         \label{MONSDynM}
\begin{tabular}{crr}
\noalign{\hrule}
Planet& $M_h$($M_\oplus$) & $M_\alpha$($M_\oplus$) \\
\noalign{\hrule}
Mercury & 0.055 & 0.4 \\
Venus & 0.81 & 0.37 \\
Earth & 1.0 & 0.48 \\
Mars & 0.11 & 2.6(0.30) \\
Jupiter & 11-42 & 4.9(7.1) \\
Saturn & 19-31 & 8.4 \\
Uranus & 12.5 & 12.5 \\
Neptune & 15 & 12 \\
\noalign{\hrule}
\end{tabular}
   \end{table}

\section{Viscosity values in different regions of the nebula}

I use the widely accepted approach for $\alpha$ values (Papaloizou
\& Lin 1995; Stone et al. 2000; Balbus 2003). For the case of the
solar nebula, see the review by Stone et al. (2000) and references
therein. It seems that hydrodynamic turbulence is ineffective as
an AMT mechanism. Gravitational instability can transport angular
momentum when the nebula is massive (Laughlin \& Bodenheimer 1994;
Papaloizou \& Lin 1995). The effective value of $\alpha$ is $\sim
0.03-0.1$. This can dominate the AMT during the early stage. Much
of the stellar mass may be gained this way. As the nebula mass
drops, less efficient AMT processes take over. The MHD
(magnetohydrodynamic) turbulence driven by the magnetorotational
instability (MRI) is a very likely mechanism (Stone et al. 2000).
The viscosity is high (low) when the MRI can (not) survive. The
$ideal$ MHD simulations give values of $\alpha$ ranging from $5
\times 10^{-3}$ to $\sim 0.6$. The high value is reached when
there is a net vertical field. A typical value for $\alpha$ is
10$^{-2}$. The solar magnetic field may provide such a net
vertical field inside Mercury if the solar dynamo starts that
early. Wave propagation alone is a less effective AMT than the MHD
turbulence. The excitation is most powerful in the outer region of
the nebula. This may favor high $\alpha$ in the outer region. The
value used to fit observations of accretion rates (Hartmann et al.
1998) is $\alpha \sim 10^{-2}$. The age consideration also
indicates $\alpha \sim 10^{-3}-10^{-2}$.

   \begin{figure}
   \centering
   \rotatebox{0}{
   \includegraphics[width=88mm]{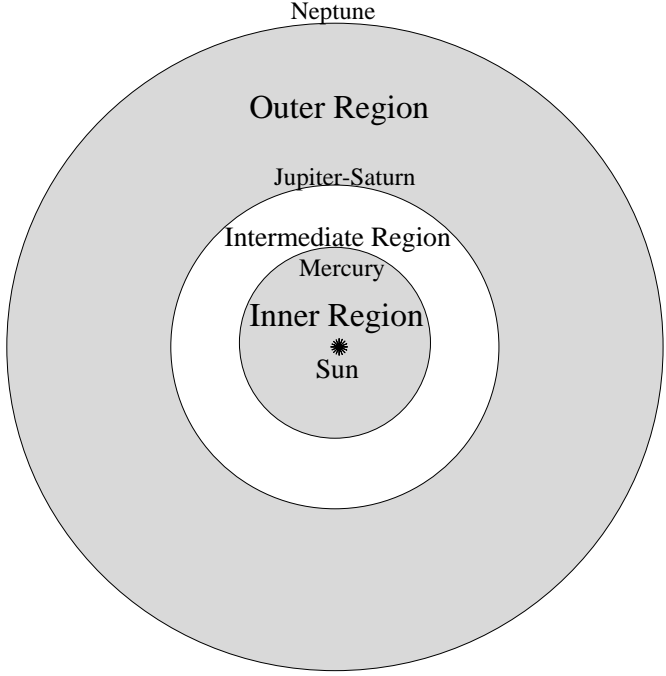}}
   \caption{
A top view of the solar nebula illustrating the various regions of
the viscosity. {\bf Inner Region}: MRI (magnetorotational
instability) survives due to the thermal ionization.High
viscosity, high radial inflow velocity, and short inflow time.
{\bf Intermediate Region}: MRI is suppressed. Low viscosity, low
inflow velocity, and long inflow time. {\bf Outer Region}: MRI
survives due to the ionization by cosmic rays. High viscosity,
high inflow velocity, and short inflow time. The material tends to
accumulate in the Jupiter-Saturn region due to the difference of
the inflow velocity between the intermediate and outer regions.
The gas inflow time in Jupiter-Saturn region is long enough that
they have time to capture the gas. }
              \label{FigHR}%
    \end{figure}

Jin (1996) considered the effect of ohmic diffusion on the MRI and
showed that the MRI is damped when the diffusion rate is greater
than the MRI growth rate. The diffusion rate is high (low) when the
ionization degree is low (high). The MRI can survive at temperature
$T>1000$K due to the thermal ionization. This temperature can be
reached in the inner region of the nebula inside Mercury. Cosmic
rays can partially ionize the part of the nebula where they can
penetrate (Hayashi 1981). So the cosmic ray ionization is more
significant where $\Sigma$ is low, which is the outer region of the
nebula. Thus the MRI can survive there. An estimate of the location
of the transition zone between the outer region and the intermediate
region (where the MRI can not survive) can be found by equating the
cosmic ray penetration depth $\Sigma_{CR}$ = 100 g cm$^{-2}$ with
$\Sigma$. For an easy estimate, I use Hayashi (1981) surface density
and find that the transition zone is around $r \sim$7AU, which is
Jupiter-Saturn region. Notice that this radius is larger when the
nebula has more mass than the minimum mass. MHD simulations (Fleming
\& Stone 2003) finds that the viscosity can drop below $\alpha \sim
10^{-4}$ where $\Sigma$ is large. To summarize, {\it the MHD
turbulence driven by the MRI causes high $\alpha$ $(\sim 10^{-2})$
in the inner region inside Mercury and the outer region outside
Jupiter-Saturn, and $\alpha$ $(\sim 10^{-4})$ is significantly lower
in the intermediate region due to the damping of the MRI} (Fig. 1).

\section{Mass accumulation and gas trapping of Jupiter and Saturn}

Lets look at the mass enhancement in Jupiter-Saturn region due to
the above nonuniform $\alpha$ (AMT). The radial inflow velocity is
(Pringle 1981)
\begin{equation}
v_r \approx -{\nu \over r} = -\alpha c_s {H \over r} = -\alpha
\left({H \over r}\right)^2 r\Omega,
\end{equation}
where $\nu$ is the kinematic viscosity, $c_s$ is the sound speed,
$H$ is the nebula thickness, and $\Omega$ is the angular velocity.
The negative sign means that the material flows inwards to the
Sun. Notice that the more efficient AMT drives faster radial
inflow. {\it Therefore the inflow in the outer region is faster
than the intermediate region and the material tends to accumulate
in the transition zone} (Fig 1). This means that Jupiter and
Saturn have access to more material. Mathematically, the mass
enhancement in Jupiter-Saturn region can be shown as the
following. The mass inflow rate at any radius $r$ is (Pringle
1981)
\begin{equation}
\dot M = -2 \pi r \Sigma v_r.
\end{equation}
Let $\Sigma_0(r)$ be the solution of the surface density with
constant $\alpha =\alpha_o$ (the $\alpha$ value in the outer
region). Assuming that $\Sigma_0$ is not changed for the first
order, the mass enhancement rate in the transition zone compared
with constant $\alpha$ solution is
\begin{equation}
\begin{array}{rl}
\Delta \dot M = (2 \pi r \Sigma_0 v_r |_{r=R_i} - 2 \pi r \Sigma_0
v_r |_{r=R_o})_{\alpha = \alpha_n} \\
-(2 \pi r \Sigma_0 v_r
|_{r=R_i} - 2 \pi r \Sigma_0 v_r |_{r=R_o})_{\alpha = \alpha_o},
\end{array}
\end{equation}
where $R_i$ and $R_o$ are the inner and outer radii of the
transition zone, and $\alpha_n$ is the nonuniform $\alpha$. The
term in first parenthesis is the mass change rate in the zone with
nonuniform $\alpha$ and the second is the rate  with constant
$\alpha=\alpha_o$. Since $\alpha_n = \alpha_o$ at $r=R_o$, two
terms at $R_o$ are cancelled. We have by using equation (1)
\begin{equation}
\Delta \dot M = 2 \pi r \Sigma_0 (\alpha_o-\alpha_i) \left({H
\over r}\right)^2 r\Omega |_{r=R_i},
\end{equation}
where $\alpha_i$ is the $\alpha$ value in the intermediate region.
It is straightforward to see that there is a mass enhancement in
the transition zone which is Jupiter-Saturn region since
$\alpha_i$ is significantly lower than $\alpha_o$. The enhancement
is due to the difference in the radial velocity caused by the
difference of $\alpha$ value. If the nebula ever reaches a
quasi-steady state ($\dot M$ does not change with $r$), equation
(2) gives that $\Sigma$ is higher in the  lower $\alpha$ region
which is the intermediate region.

In the process of the planet formation, grains are decoupled from
the gas when they grow into larger solid bodies. The inflow keeps
the same initial solar composition before the decoupling. {\it I
suggest that the mass enhancement in Jupiter-Saturn region before
the decoupling explains their enhancement of heavy element masses
(Table 1)}. After the decoupling, the gas will continue its
inflow. The inflow time is (Pringle 1981)
\begin{equation}
t_\nu \approx -{r \over v_r}  \approx \alpha^{-1} \left({H \over
r}\right)^{-2} \Omega^{-1} = 72  \alpha^{-1} \left({r\over
AU}\right) yr,
\end{equation}
where the value of $H/r$ from Hayashi (1981) is used. I list
$\alpha$ and calculated inflow times at the heliocentric radius of
the planets in Table 2. After formation of solid cores, the first
phase of the gas trapping of giant planets is the slow capture of
the surrounding hydrogen-helium which takes a few $10^6$yr. Then a
rapid accretion of the gas only takes $\sim 10^5$yr (Pollack et
al. 1996). By comparing the slow capture time with the inflow time
(Table 2), {\it the main discovery of this letter is the
following}. The inflow is faster than the capture for Uranus and
Neptune and slower for Jupiter and Saturn. Therefore before Uranus
and Neptune can capture enough gas to reach the rapid accretion,
the gas already flows to Jupiter-Saturn region. This explains the
low masses and low H-He abundances of Uranus and Neptune among the
Jovian planets. But Jupiter and Saturn have enough time to capture
the gas before the gas can flow further inward so that they have
large masses and large H-He abundances (Fig 1). The fact that
Jupiter has bigger mass than Saturn is rarely addressed. I suggest
that because Jupiter region has lower $\alpha$ (the higher
$\Sigma$, the less cosmic ray penetration) than Saturn, Jupiter
has more time to trap gas.

   \begin{table}
   \centering
      \caption[]{Inflow time (in year) at the heliocentric radius $r$ of the planets.}
         \label{MONSDynM}
\begin{tabular}{crrr}
\noalign{\hrule}
Planet& $r$(AU) & $\alpha$ & $t_\nu$(yr) \\
\noalign{\hrule}
Mercury & 0.39 & $10^{-2}$ & $3 \times 10^3$ \\
Venus & 0.72 & $10^{-4}$ & $5 \times 10^5$ \\
Earth & 1.0 & $10^{-4}$ & $7 \times 10^5$ \\
Mars & 1.5 & $10^{-4}$ & $1 \times 10^6$ \\
Jupiter & 5.2 & $10^{-4}$ & $4 \times 10^6$ \\
Saturn & 9.5 & $10^{-4}$ & $7 \times 10^6$ \\
Uranus & 19 & $10^{-2}$ & $1 \times 10^5$ \\
Neptune & 30 & $10^{-2}$ & $2 \times 10^5$ \\
\noalign{\hrule}
\end{tabular}
   \end{table}

\section{Application to other planets and discussion}

Mars is the terrestrial planet adjacent to the Jovian planets. The
famous Mars drop (Table 1) might be caused by the sweeping of
early-formed Jupiter. If the sweeping extends to Mars orbit,
calculated masses for Jupiter and Mars with $\Sigma \sim r^{-1}$
are put in parenthesis in Table 1. The match with observed masses
are much better for Mars. The mass enhancement of Earth and Venus
(Table 1) is due to the low viscosity (slow inflow). AMT is very
efficient in the inner region inside Mercury because of MHD
turbulence and it might be even enhanced due to the possible net
flux of the solar magnetic field. In addition, magnetic winds may
contribute to AMT (Papaloizou \& Lin 1995). Contrary to
Jupiter-Saturn region, the material tends to deplete in Mercury
region because the inflow is faster in the inner region than in
the intermediate region (Fig 1). So this theory also provides a
natural interpretation of the low mass of Mercury (Table 1).
Notice that this theory does not contradicts theory of dynamics of
planetesimals.

Some of advantages of the new theory presented in this letter are:
1) It is simple and comes natural because it is, without any
additional contrived assumptions or physical processes (therefore
no new uncertainties), based on the well known wisdom that there
is AMT during the nebula evolution in order to understand the
current distribution of the angular momentum of the Solar System.
The mass distribution due to the AMT is $inevitable$. 2) The
theory uses only one physical mechanism, AMT, to give an unified
picture of the planet masses and compositions. In addition to the
large masses of Jupiter and Saturn and their large amount of H and
He, the theory might explain low mass of Mercury, the difference
between Saturn and Jupiter, and the enhancement of heavy elements
of Saturn, Jupiter, Earth, and Venus relative to the model with
constant $\alpha$. These facts are rarely addressed. No previous
theories have put all of these together. 3) The interpretation for
the difference between Jupiter-Saturn and Uranus-Neptune does not
depend on the details of the formation processes of the cores of
the Jovian planets. The mass redistribution due to the nonuniform
AMT is independent of details of planet formation. This theory
does not in any way contradicts existing models of planet
formation. This would make the theory more viable. Although
researchers can have different interpretations for planet masses,
it seems clear to the author that the impact of the nonuniform AMT
on the planet masses can not be ignored. Notice that the results
of this letter depend only on the final values of viscosity, but
not on details of AMT, such as what drives AMT.

\begin{acknowledgements}
I thank Donald Clayton for reading and comments on the present
work and Jin Wang for suggestions of revising the manuscript.
\end{acknowledgements}

\end{document}